# Preliminary heavy-light decay constants from the MILC collaboration[*]

C. Bernard,[a] T. Blum,[b] A. De,[a] T. DeGrand,[c] C. DeTar,[d] Steven Gottlieb,[e] U.M. Heller,[f] N. Ishizuka,[a] L. Kärkkäinen,[b] J. Labrenz,[g] A. Soni,[h] R. Sugar,[i] and D. Toussaint[b]

[a]Department of Physics, Washington University, St. Louis, MO 63130, USA

[b]Department of Physics, University of Arizona, Tucson, AZ 85721, USA

[c]Physics Department, University of Colorado, Boulder, CO 80309, USA

[d]Physics Department, University of Utah, Salt Lake City, UT 84112, USA

[e]Department of Physics, Indiana University, Bloomington, IN 47405, USA

[f]SCRI, The Florida State University, Tallahassee, FL 32306-4052, USA

[g]Physics Department, University of Washington, Seattle, WA 98195, USA

[h]Physics Department, Brookhaven National Laboratory, Upton, NY 11973, USA

[i]Department of Physics, University of California, Santa Barbara, CA 93106, USA

Preliminary results from the MILC collaboration for $f_B$, $f_{B_s}$, $f_D$, $f_{D_s}$ and their ratios are presented. We compute in the quenched approximation at $\beta = 6.3$, 6.0 and 5.7 with Wilson light quarks and static and Wilson heavy quarks. We attempt to quantify all systematic errors other than quenching.

## 1. PRELIMINARIES

Over the past year, we have been computing heavy-light decay constants in the quenched approximation on Intel Paragon computers. Most of the computations have been performed on the 512-node Paragon at Oak Ridge National Laboratory, but Paragons at Indiana University and at the San Diego Supercomputer Center have also been used. In many respects the calculations are standard, and we emphasize here only the distinguishing features.

The initially very slow I/O speeds of the Paragon and the lack of long-term storage capability at ORNL forced us to do all the computations "on the fly." The hopping parameter computation of the heavy quark propagator, as suggested by Henty and Kenway [1], makes such on-the-fly computations possible for heavy-light systems. Because the full light and heavy propagators for all spin-color sources can not be stored in memory, we work only with one spin-color source for light and heavy at a time, and restrict ourselves to mesons which are diagonal in spin-color (*i.e.*, pseudoscalars and the z-component of vectors). We run 400 hopping parameter passes. At $\beta = 6.3$, this gives very good convergence for heavy quarks with $\kappa_h \leq 0.145$.

Gaussian quark sources in Coulomb gauge are used. The overrelaxed gauge fixer is run until the sum of the trace of all spacelike links (normalized to 1 when all links are unit matrices) changes by less than $7 \times 10^{-7}$ per pass. On the $24^3 \times 80$ lattices at $\beta = 6.3$, this takes about 435 passes. The half-width of the gaussian is $\approx 0.4$ fm.

Table 1
Lattice parameters.

| name | $\beta$ | size | # configs. (planned) |
|---|---|---|---|
| A | 5.7 | $8^3 \times 48$ | 200 (200) |
| B | 5.7 | $16^3 \times 48$ | 100 (100) |
| C | 6.0 | $16^3 \times 48$ | 48 (100) |
| D | 6.3 | $24^3 \times 80$ | 98 (100) |

---

[*]presented by C. Bernard

We compute "smeared-local" and "smeared-smeared" propagators. Because the mesons must be constructed at each of the 400 orders of the hopping parameter expansion, it is too expensive to sum the central point of the smeared sinks over the entire spatial volume, even using FFT's. Instead, we simply sum over a subset of the points in the spatial volume. This allows intermediate states of non-zero 3-momentum to contribute. For the heavy-light mesons studied here, the higher momentum states are well suppressed at asymptotic time by their higher energy. However, the static-light mesons have no such suppression, and the contribution of higher momentum states is limited only by their overlap with the sources.

We sum the sinks over 16 points on a time slice. Using computed static-light wavefunctions [2], we find that the contamination in static-light decay constants from nonzero momentum states is small ($\approx 0.7\%$) for lattices with spatial size of $\approx 1.5$ fm (lattices A, C, and D). However, on lattice B, with spatial size of $\approx 3$ fm, the contamination is $\approx 60\%$. We thus do not include the static point from lattice B (nor from lattice A, so we may compare the $\beta = 5.7$ lattices without bias).

Since we only have results for degenerate light quarks, we determine $\kappa_s$, the strange quark hopping parameter, by adjusting the pseudoscalar mass to $\sqrt{2m_K^2 - m_\pi^2}$, the lowest order chiral perturbation theory value.

For heavy-light mesons we use the Kronfeld-Mackenzie [3] norm ($\sqrt{1 - 6\tilde{\kappa}}$) and adjust the measured meson pole mass upward by the difference between the heavy quark pole mass ("$m_1$") and the heavy quark dynamical mass ("$m_2$") as calculated in the tadpole-improved tree approximation [3].

## 2. RESULTS

A plot of $f_P\sqrt{M_P}$ vs. $1/M_P$ is shown in fig. 1 for lattice D. The fit is covariant, to the form $c_0 + c_1/M_P + c_2/M_P^2$. The $\chi^2$/d.o.f for the fit is $\approx 2$ (confidence level $\approx 10\%$), whether or not the static-light point is included. The rather low con-fidence level may perhaps be due to the fact that we have not included additional large-$ma$ corrections to the action and operators [4], or simply to the small differences between the heavy quark mass and the meson mass $M_P$. Such effects are under investigation. Note that, in an earlier calculation [5], the statistical errors were considerably larger, and the $\chi^2$/d.o.f for such fits was good. Here the level of statistical precision has increased to a level where small effects are becoming important.

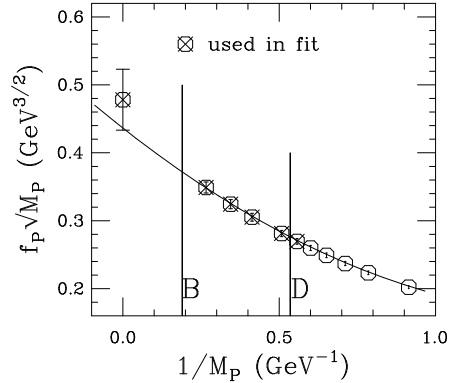

Figure 1. $f_P(M_P)^{\frac{1}{2}}$ vs. $1/M_P$ for lattice D. The light quark is extrapolated to the physical mass $(m_u + m_d)/2$.

Table 2
Results for decay constants and ratios. $f_\pi = 131$ MeV scale used throughout.

|  | A | B | C | D |
|---|---|---|---|---|
| $f_B$ | 195(6) | 198(4) | 175(5) | 166(4) |
| $f_{B_s}$ | 244(5) | 237(3) | 207(4) | 192(3) |
| $f_D$ | 227(5) | 227(4) | 205(4) | 198(2) |
| $f_{D_s}$ | 275(4) | 273(3) | 239(3) | 225(2) |
| $\frac{f_{B_s}}{f_B}$ | 1.25(2) | 1.20(1) | 1.18(1) | 1.16(1) |
| $\frac{f_{D_s}}{f_D}$ | 1.21(1) | 1.20(1) | 1.17(1) | 1.13(1) |

Table 2 shows results from the four lattices. The lattice-spacing dependence is apparent, but little, if any, finite volume effect is present (compare A and B). This is seen more clearly in Fig. 2, which shows $f_B$ vs. lattice spacing. It is natural to extrapolate the $f_\pi$-scale results linearly to the continuum; we get 147(6) MeV. Note that the $f_\pi$-scale results have much less $a$ dependence than those using $m_\rho$. This makes sense since $f_\pi$ and $f_B$ are likely to have rather similar finite $a$ effects. If the $m_\rho$-scale results are extrapolated linearly to $a = 0$, the result is 119(8) MeV, considerably smaller than the $f_\pi$-scale extrapolation. However, there also seem to be larger finite volume effects in the $m_\rho$-scale results, which is reasonable since the $\rho$ is a larger state than the $\pi$. If we first adjust upward the results from lattices C and D by the presumed finite volume correction obtained by comparing the lattices B and A, the result of the $m_\rho$-scale extrapolation (144(9) MeV) is consistent with the $f_\pi$-scale result.

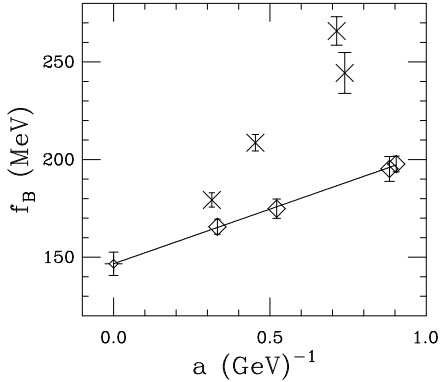

Figure 2. $f_B$ vs. $a$. Diamonds have scale set by $f_\pi$; crosses, by $m_\rho$. Fit is to the diamonds; "fancy diamond" gives the $a = 0$ extrapolation. The higher cross at $a \approx 0.7$ is from lattice B.

The results on lattice D are consistent with those of [5]. The major cause of the somewhat smaller central values here is a lower (but still consistent) estimate of the scale ($1/a \approx 3.0$ GeV here vs. $\approx 3.2$ GeV in [5]).

We linearly extrapolate to $a = 0$ all results in Table 2. Systematic errors are then estimated — in a very preliminary fashion — as follows: (1) Changes of fitting ranges (in $t$) for the propagators and of types of fits in $1/M$ for $f_P\sqrt{M_P}$ give a typical variation of about twice the statistical errors. (2) The dependence on the determination of $\kappa_s$ is estimated by finding the change in the extrapolated results if $\kappa_s$ is fixed instead using the vector state $\phi$. The difference is especially significant for $f_{B_s}/f_B$ and is $\approx 0.05$ there. (3) Finite volume effects are estimated by taking the fractional difference between results from lattices A and B, using the $m_\rho$ scale. This is conservative, since the central values are found with the $f_\pi$ scale. (4) We estimate scale errors by comparing the results at $\beta = 6.3$ with $f_\pi$ and $m_\rho$ scales. The difference ($\approx 13$ MeV for the decay constants and $\approx 0.02$ for the ratios) is roughly comparable to what we would get by comparing extrapolated $f_\pi$- and $m_\rho$-scale values, after adjusting for the apparent finite volume effects. (5) The effects of using heavy Wilson fermions without the additional corrections to the action and operators detailed in [3,4] are estimated by comparing the original fits at $\beta = 6.3$ (see, e.g., Fig. 1) with fits using only the 6 lightest heavy-light states (and the static-light point). In the original fits the maximum value of $(m_2 - m_1)/m_2$ is 0.22; in the new ones, 0.04. The differences in the results are quite small: $\approx 4$ MeV for the decay constants and $\approx 0.01$ for the ratios.

Adding the above systematic errors in quadrature, our preliminary results are

$$f_B = 147(6)(23); \quad f_D = 181(4)(18);; \quad (1)$$
$$f_{B_s} = 164(5)(20); \quad f_{D_s} = 195(3)(16); \quad (2)$$
$$\frac{f_{B_s}}{f_B} = 1.13(2)(8); \quad \frac{f_{D_s}}{f_D} = 1.09(1)(4), \quad (3)$$

where the decay constants are in MeV. Study of the quenching errors is in progress.

We thank A. El-Khadra, A. Kronfeld, and P. Mackenzie for useful conversations. Computing was done at ORNL Center for Computational Sciences, SDSC, and Indiana University. This work was supported in part by the DOE and NSF.